\documentclass[aps,prd,nofootinbib,twocolumn,preprintnumbers,superscriptaddress,showpacs]{revtex4-1}

\usepackage{amsmath}
\usepackage{amsthm}
\usepackage{amssymb}
\usepackage{graphicx}
\usepackage{bm}

\newcommand{\beq}{\begin{equation}}
\newcommand{\eeq}{\end{equation}}
\newcommand{\beqa}{\begin{eqnarray}}
\newcommand{\eeqa}{\end{eqnarray}}

\begin{document}

\title{General Wahlquist metrics in all dimensions}

\author{Kazuki Hinoue}
\email{hinoue@sci.osaka-cu.ac.jp}
\affiliation{Department of Mathematics and Physics, Graduate School of Science,
Osaka City University, Osaka 558-8585, Japan}

\author{Tsuyoshi Houri}
\email{houri@rikkyo.ac.jp}
\affiliation{Department of Physics, Rikkyo University, Tokyo 171-8501, Japan}

\author{Christina Rugina}
\email{cristina.rugina11@imperial.ac.uk}
\affiliation{Department of Physics, University of Bucharest, Bucharest 050107, Romania,}
\affiliation{IFIN-HH, Department of Theoretical Physics, Magurele 07712, Romania}
\affiliation{Department of Physics, Imperial College, London SW7 2AZ, United Kingdom}

\author{Yukinori Yasui}
\email{yasui@sci.osaka-cu.ac.jp}
\affiliation{Department of Mathematics and Physics, Graduate School of Science,
Osaka City University, Osaka 558-8585, Japan}

\date{\today}

\begin{abstract}
It is shown that the Wahlquist metric, which is a stationary,
axially symmetric perfect fluid solution with $\rho+3p=\text{const}$,
admits a rank-2 generalized closed conformal Killing-Yano tensor with a skew-symmetric torsion.
Taking advantage of the presence of such a tensor, 
we obtain a higher-dimensional generalization of the Wahlquist metric in arbitrary dimensions,
including a family of vacuum black hole solutions with spherical horizon topology
such as Schwarzschild-Tangherlini, Myers-Perry and higher-dimensional Kerr-NUT-(A)dS metrics
and a family of static, spherically symmetric perfect fluid solutions in higher dimensions.
\end{abstract}

\pacs{02.40.Ky, 04.20.Jb, 04.50.Gh}

\preprint{OCU-PHYS 398}
\preprint{RUP-14-3}

\maketitle

\section{Introduction}
Since the discovery of the Kerr metric which describes rotating black holes in a vacuum,
its geometry has been investigated from the viewpoint of classifying spacetimes
to understand what are the most fundamental properties of the Kerr spacetime.
A number of studies for the purpose have been conducted
in various systematical frameworks (e.g., see \cite{Stphani:2003}),
and we have obtained a common understanding as to the Kerr spacetime to date:
stationary, axially symmetric, asymptotically flat,
Petrov type D vacuum solution of the vanishing of the Simon tensor,
admitting a rank-2 Killing-St\"ackel (KS) tensor of Segre type $[(11)(11)]$ 
constructed from a (nondegenerate) rank-2 Killing-Yano (KY) tensor.

The Wahlquist metric \cite{Wahlquist:1968,Kramer:1985,Senovilla:1987,Wahlquist:1992,Mars:2001}
investigated in this paper
was found in the study of stationary, axially symmetric perfect fluid spacetimes.
As we succeeded in obtaining interior solutions of static, spherically symmetric stars
joined to the Schwarzschild vacuum spacetime,
it has been thought that the interior of rotating bodies can be described
by stationary, axially symmetric perfect fluid solutions.
Although it was shown that the Wahlquist metric cannot be smoothly matched
to an asymptotically flat vacuum spacetime \cite{Bradley:1999},
the metric is still of great interest because
it allows some interesting geometric characterizations.
It was pointed out \cite{Kramer:1985} that
the Wahlquist metric is the general solution of
stationary, rigidly rotating perfect fluids
with the vanishing Simon tensor.
It was also demonstrated \cite{Senovilla:1987} that the metric
is the general solution of Petrov type D, stationary, axially symmetric
and rigidly rotating perfect fluids with $\rho+3p=\text{const}$.
Furthermore, the Wahlquist metric is known to be included in the class of metrics
admitting a rank-2 KS tensor of Segre type $[(11)(11)]$
which possesses two double nonconstant eigenvalues \cite{HM1,HM2}.

The Wahlquist spacetime inherits some geometric properties of the Kerr spacetime.
This seems to be reasonable because
the Kerr metric is obtained as the limiting case of the Wahlquist metric \cite{Wahlquist:1968,Mars:2001}.
In the Kerr spacetime, two Killing vectors and KS tensor
are constructed from a single rank-2 KY tensor.
It is also shown that the Kerr metric is the only asymptotically flat vacuum solution
admitting a rank-2 KY tensor \cite{Dietz:1982}.
This implies that the presence of the KY tensor is essential to characterize the Kerr spacetime.
Nevertheless, Killing-Yano symmetry of the Wahlquist spacetime has never been investigated.
In this paper, therefore, we first ask whether the Wahlquist metric admits Killing-Yano symmetry.
Actually, as we will see in Sec.~II,
we find a rank-2 generalized closed conformal Killing-Yano (GCCKY) tensor with torsion \cite{KKY}
for the Wahlquist metric.

In recent years, it has been unveiled that Killing-Yano symmetry plays an important role
in higher-dimensional rotating black hole spacetimes.
A family of vacuum solutions describing rotating black holes with spherical horizon topology
\cite{Myers:1986,Hawking:1999,Gibbons:2004,Chen:2006}
admits a rank-2 closed conformal Killing-Yano (CCKY) tensor \cite{Houri:2007,Krtous:2008,Houri:2009}.
Recently, local metrics admitting a rank-2 GCCKY tensor were classified
into three types (called type A, B and C) in arbitrary dimensions \cite{HKWY}.
Some supergravity black hole solutions in higher dimensions are included
in type A of the classification \cite{KKY,HKWY:2010}.
In this paper, by employing the classification,
we attempt to obtain a new family of rotating perfect fluid solutions
which generalizes the Wahlquist solution to higher dimensions.
On the other hand, there have been static, spherically symmetric perfect fluid solutions
in higher dimensions \cite{Krori:1988,Shen:1989,Leon:2002,Zarro:2009}.
The solutions obtained in this paper cover, in the static limit,
the static, spherically symmetric perfect fluid solutions with $\rho+3p=\text{const}$.

The paper is organized as follows.
In Sec.~II, after we briefly review the Wahlquist solution,
we demonstrate that the Wahlquist metric admits a rank-2 GCCKY tensor with a skew-symmetric torsion.
With respect to such a tensor,
in Sec.~III, we generalize the Wahlquist solution in four dimensions
to higher-dimensional ones by solving the Einstein equations for perfect fluids
in all even and odd dimensions, respectively.
We see that the equations of state for the higher-dimensional Wahlquist solutions
are given by $\rho+3p=\text{const}$ in all dimensions.
In Sec.~IV, we investigate the separability of the Hamilton-Jacobi for geodesics,
Klein-Gordon and Dirac equations in all dimensions.
Finally, Sec.~V is devoted to summary and discussion.
After we review the geometry of the four-dimensional Wahlquist spacetime in Appendix A,
we discuss the five-dimensional Wahlquist metric in Appendix B.
In Appendix C, we have collected the curvature quantities of the higher-dimensional Wahlquist metrics.
As a result, it is shown that the metrics are of type D in the higher-dimensional Petrov classification \cite{Coley:2004}.

\section{Killing-Yano Symmetry of the Wahlquist Spacetime}
To investigate Killing-Yano symmetry of the Wahlquist metric in four dimensions
\cite{Wahlquist:1968,Kramer:1985,Senovilla:1987,Wahlquist:1992,Mars:2001},
we begin with the metric form that appeared in \cite{Mars:2001},
which is written in a local coordinate system ($z,w,\tau,\sigma$) as
\beqa
ds^2 &=& (v_1+v_2)\left( \frac{dz^2}{U}+\frac{dw^2}{V} \right) \label{Wahlquist} \\
     & &  +\frac{U}{v_1+v_2}(d\tau+v_2 d\sigma)^2-\frac{V}{v_1+v_2}(d\tau-v_1 d\sigma)^2 \,, \nonumber
\eeqa
where
\beqa
 U &=& Q_0 + a_1 \frac{\sinh(2\beta z)}{2\beta} - \nu_0\frac{\cosh(2\beta z)-1}{2\beta^2} \nonumber\\
   &&  - \frac{\mu_0}{\beta^2}\left[\frac{\cosh(2\beta z)-1}{2\beta^2}-\frac{z\sinh(2\beta z)}{2\beta}\right] \,, \nonumber\\
 V &=& Q_0 + a_2 \frac{\sin(2\beta w)}{2\beta} + \nu_0\frac{1-\cos(2\beta w)}{2\beta^2} \\
   &&  + \frac{\mu_0}{\beta^2}\left[\frac{1-\cos(2\beta w)}{2\beta^2}-\frac{w\sin(2\beta w)}{2\beta}\right] \,, \nonumber
\eeqa
and
\begin{equation}
 v_1 = \frac{\cosh(2 \beta z)-1}{2 \beta^2} \,, \qquad
 v_2 = \frac{1-\cos(2 \beta w)}{2 \beta^2} \,.
\end{equation}
The metric contains six real constants $Q_0$, $a_1$, $a_2$, $\nu_0$, $\mu_0$ and $\beta$.
Since one of them can be eliminated by coordinate transformation,
only five of the constants are independent.
As was shown in \cite{Wahlquist:1968,Mars:2001}, one can take the limit $\beta\to 0$,
in which the metric reduces to the Kerr-NUT-(A)dS metric \cite{Carter:1968}
(see Appendix A for details).

The Wahlquist metric provides the stress-energy tensor for perfect fluids
of the energy density $\rho$, pressure $p$ and 4-velocity ${\bm u}$ with $u_\mu u^\mu=-1$,
which is written as
\beq
 T^{\mu\nu} = (\rho+p)u^\mu u^\nu + p g^{\mu\nu} \,.
\eeq
The 4-velocity is given by
\beq
 u^\mu \frac{\partial}{\partial x^\mu} = \frac{1}{\sqrt{-g_{\tau\tau}}}\frac{\partial}{\partial \tau} \,,
\eeq
where $g_{\tau\tau}=(U-V)/(v_1+v_2)$.
When we consider stationary, axially symmetric spacetimes,
we have two Killing vector fields ${\bm \partial}_t$ and ${\bm \partial}_\phi$.
If ${\bm u}$ lies on the 2-plane spanned by the two Killing vector fields,
then ${\bm u}$ can be written as ${\bm u} = N({\bm \partial}_t+\Omega{\bm \partial}_\phi)$
where $N$ and $\Omega$ are functions in general.
In particular, when $\Omega$ is constant,
the perfect fluid is said to be rigidly rotating.
Namely, the Wahlquist solution represents rigidly rotating perfect fluids.
The energy density and pressure are given by
\beq
 \rho = -\mu_0 - 3\beta^2 g_{\tau\tau} \,, \qquad
 p = \mu_0+\beta^2g_{\tau\tau} \,.
\eeq
Thus, the equation of state is $\rho+3p=2\mu_0$.
Since we have $\rho+p=0$ and $p=\mu_0$ in the limit $\beta \to 0$,
the constant $\mu_0$ is the cosmological constant.

\subsection{Generalized Killing-Yano symmetry}
It is known that the Kerr metric admits separation of variables in the Hamilton-Jacobi for geodesics,
Klein-Gordon and Dirac equations.
The separability is due to the presence of a rank-2 KY tensor.
In four dimensions, the Hodge dual of the KY tensor is a
rank-2 CCKY tensor ${\bm h}$ \cite{Tachibana:1969} satisfying
\beq
 \nabla_ah_{bc}  = g_{ab}\xi_c - g_{ac}\xi_b \,, \label{KY}
\eeq
where $\nabla$ is the Levi-Civit\`a connection.
From (\ref{KY}), the associated vector ${\bm \xi}$ is obtained as
\beq
 \xi_a = \frac{1}{3} \nabla^bh_{ba} \,.
\eeq
Namely, the Kerr metric admits a rank-2 CCKY tensor.

The Wahlquist metric partially shares the separability of the Kerr metric:
the Hamilton-Jaocbi for geodesics and Klein-Gordon equations separate,
but the Dirac equation does not.
Since the Kerr metric is obtained as a particular limit of the Wahlquist metric \cite{Wahlquist:1968,Mars:2001},
it is natural to ask the Wahlquist metric to admit a rank-2 CCKY tensor.
However, it is shown that such a tensor does not exist in the Wahlquist spacetime.
Instead, we find a rank-2 GCCKY tensor ${\bm h}$ \cite{KKY},
with a skew-symmetric torsion ${\bm T}$ satisfying
\beq
 \nabla^T_ah_{bc} = g_{ab}\xi_c - g_{ac}\xi_b \,, \label{GCCKYeq}
\eeq
where $\nabla^T$ is the connection with the skew-symmetric torsion defined by
\beq
 \nabla^T_ah_{bc} = \nabla_ah_{bc} + \frac{1}{2} T_{ab}{}^dh_{cd} - \frac{1}{2} T_{ac}{}^d h_{bd} \,.
\eeq
The associated vector ${\bm \xi}$ is given by
\beq
 \xi_a = \frac{1}{3}\nabla^T{}^bh_{ba} \,.
\eeq
If a rank-2 GCCKY tensor is obtained, we may expect
that a modified Dirac equation with $1/3$ torsion separates \cite{HKWY2}.
In fact, the modified Dirac equation of the Wahlquist metric does.
Thus, the GCCKY tensor underpins the separability
on the Hamilton-Jacobi for geodesics, Klein-Gordon and modified Dirac 
equations of the Wahlquist metric (see Sec.~IV for details).

Going through the following steps, 
we demonstrate that the Wahlquist metric (\ref{Wahlquist}) admits
a rank-2 GCCKY tensor.
To see it, we first introduce the coordinates $x$ and $y$ defined by
\begin{equation}
 x^2 = v_1 \,, \qquad y^2 = v_2 \,,
\end{equation}
and hence
\begin{equation}
 dz^2 = \frac{dx^2}{\beta^2 x^2+1} \,, \qquad
 dw^2 = \frac{dy^2}{1-\beta^2 y^2} \,.
\end{equation}
The metric is then written as
\beqa
ds^2 &=& \frac{x^2+y^2}{U(1+\beta^2 x^2)}dx^2+\frac{x^2+y^2}{V(1-\beta^2 y^2)}dy^2 \label{Wah} \\
     & & +\frac{U}{x^2+y^2}(d\tau+y^2 d\sigma)^2-\frac{V}{x^2+y^2}(d\tau-x^2 d\sigma)^2 \nonumber
\eeqa
with the functions
\beqa
 U &=& Q_0 + a_1 x\sqrt{1+\beta^2x^2} - \nu_0 x^2 \nonumber\\
   &&    - \frac{\mu_0}{\beta^2}\left[x^2-\frac{x \text{Arcsinh}(\beta x) \sqrt{1+\beta^2x^2}}{\beta}\right] \,, \nonumber\\
 V &=& Q_0 + a_2 y\sqrt{1-\beta^2y^2} + \nu_0 y^2 \\
   &&    + \frac{\mu_0}{\beta^2}\left[y^2-\frac{y \text{Arcsin}(\beta y)\sqrt{1-\beta^2y^2}}{\beta}\right] \,. \nonumber
\eeqa
Furthermore, taking the Wick rotation $y \rightarrow \sqrt{-1}y$
(with $a_2\rightarrow -\sqrt{-1}a_2$ to keep the metric function $V$ real)
and changing the sign $\sigma \rightarrow -\sigma$,
we obtain the Euclidean expression, in which the metric takes a symmetric form with respect to the coordinates $(x,y)$ as
\beqa
 ds^2_E &=& \frac{f_1(x^2-y^2)}{\Xi_1}dx^2 + \frac{f_2(y^2-x^2)}{\Xi_2}dy^2 \label{Euclidean_metric} \\
     & & + \frac{\Xi_1}{x^2-y^2}(d\tau+y^2 d\sigma)^2 + \frac{\Xi_2}{y^2-x^2}(d\tau+x^2 d\sigma)^2 \,, \nonumber
\eeqa
where
\beqa
 f_1 &=& \frac{1}{\sqrt{1+\beta^2 x^2}} \,, \quad
 f_2 = \frac{1}{\sqrt{1+\beta^2 y^2}} \,, \nonumber\\
 \Xi_1 &=& Q_0 + a_1 x\sqrt{1+\beta^2x^2} - \nu_0 x^2 \nonumber\\
   & & - \frac{\mu_0}{\beta^2}\left[x^2-\frac{x \text{Arcsinh}(\beta x) \sqrt{1+\beta^2x^2}}{\beta}\right] \,, \label{metfunc3} \\
 \Xi_2 &=& Q_0 + a_2 y\sqrt{1+\beta^2y^2} - \nu_0 y^2 \nonumber\\
   & & - \frac{\mu_0}{\beta^2}\left[y^2-\frac{y \text{Arcsinh}(\beta y)\sqrt{1+\beta^2y^2}}{\beta}\right] \,. \nonumber
\eeqa
The form of the metric (\ref{Euclidean_metric})
precisely fits into type A of the classification in \cite{HKWY};
that is, the Wahlquist spacetime admits a rank-2 GCCKY tensor.
In fact, if we introduce an orthonormal frame
\begin{eqnarray}\label{orthvector}
&& {\bm e}^1 = f_1 \sqrt{\frac{x^2-y^2}{\Xi_1}}dx \,, \qquad
   {\bm e}^2 = f_2 \sqrt{\frac{y^2-x^2}{\Xi_2}}dy \,, \nonumber\\
&& {\bm e}^{\hat{1}} = \sqrt{\frac{\Xi_1}{x^2-y^2}}(d\tau+y^2 d\sigma) \,, \\
&& {\bm e}^{\hat{2}} = \sqrt{\frac{\Xi_2}{y^2-x^2}}(d\tau+x^2 d\sigma) \,, \nonumber
\end{eqnarray}
the rank-2 GCCKY tensor is given by
\begin{equation} \label{CKY}
 {\bm h} = x \,{\bm e}^1\wedge {\bm e}^{\hat{1}} + y\,{\bm e}^2\wedge {\bm e}^{\hat{2}}
\end{equation}
with the skew-symmetric torsion
\beqa
 {\bm T} &=& \frac{2x(f_1-f_2)}{f_1f_2(x^2-y^2)} \sqrt{\frac{\Xi_2}{y^2-x^2}} 
            \,{\bm e}^1 \wedge {\bm e}^{\hat{1}} \wedge {\bm e}^{\hat{2}} \nonumber\\
     && + \frac{2y(f_2-f_1)}{f_1f_2(y^2-x^2)} \sqrt{\frac{\Xi_1}{x^2-y^2}}
           \,{\bm e}^2 \wedge {\bm e}^{\hat{2}} \wedge {\bm e}^{\hat{1}} \,. \label{torsion}
\eeqa
The torsion vanishes when we take the limit $\beta\to 0$.
This suggests that the torsion is related to the perfect fluid,
although the physical meaning of the torsion is unclear.

\section{General Wahlquist metrics in higher dimensions}
We have seen that the Wahlquist metric (\ref{Wahlquist}) admits a rank-2 GCCKY tensor
and its Euclidean form precisely fits into type A of the classification \cite{HKWY}.
Hence, it seems to be reasonable to consider a higher-dimensional generalization of the Wahlquist metric.
In this section, we attempt to solve the Einstein equations for perfect fluids in higher dimensions
by employing, as an {\it ansatz}, type A metrics in \cite{HKWY}.

We slightly change our notation
to deal with higher-dimensional metrics in both even and odd dimensions simultaneously.
We introduce $\varepsilon$
where $\varepsilon=0$ for even and $\varepsilon=1$ for odd dimensions.
The dimension number is denoted by $D=2n+\varepsilon$.
The Latin indices $a,b,\dots$ run from 1 to $D$,
and the Greece indices $\mu,\nu,\dots$ run from 1 to $n$.

The form of type A metrics in $D$ dimensions
which we deal with as an {\it ansatz}
is given by
\beqa
 {\bm g}^{(D)}
&=& \sum_{\mu=1}^n \frac{f_\mu^2}{P_\mu}dx_\mu^2
             + \sum_{\mu=1}^n P_\mu \left( \sum_{k=0}^{n-1}A^{(k)}_{\mu} d\psi_k\right)^2 \nonumber\\
& & +\varepsilon S \left(\sum_{k=0}^n A^{(k)} d\psi_k\right)^2 \,, \label{met_all}
\eeqa
where
\beqa
&& P_\mu=\frac{\Xi_\mu}{U_\mu} \,, \quad U_\mu = \prod_{\nu\neq\mu}(x_\mu^2-x_\nu^2) \,, \nonumber\\
&& S=\frac{s_0^2}{A^{(n)}} \,, \quad f_\mu=\frac{1}{\sqrt{1+\beta^2 x_\mu^2}} \,.
\eeqa
The functions $A_\mu^{(k)}$ $(k=0,\dots,n-1)$ and $A^{(k)}$ $(k=0,\dots,n)$
are $k$th order elementary symmetric functions in $\{x_1^2,x_2^2,\dots, x_n^2\}$ defined by
\beqa
&& \sum_{k=0}^{n-1}A_\mu^{(k)} t^k=\prod_{\nu \ne \mu}(1+t x_\nu^2) \,, \nonumber\\
&& \sum_{k=0}^{n}A^{(k)} t^k=\prod_{\nu=1}^n(1+t x_\nu^2) \,. 
\eeqa
The metric contains unknown functions $\Xi_\mu(x_\mu)$
depending only on single valuable $x_\mu$, and in odd dimensions a constant $s_0$.
The form of the metric (\ref{met_all}) is not the most general form of type A metrics,
but it is enough to construct a perfect fluid solution for the current purpose.
Of course, it leaves a question whether there exist more general solutions of type A.

\subsection{Tower of generalized Killing-Yano tensors}
If a rank-2 GCCKY tensor is obtained in $D=2n+\varepsilon$ dimensions,
we can construct $[D/2]=n$ conserved quantities for geodesic motion \cite{HKWY:2010}.
In addition, 
the complete integrability of the Hamilton-Jacobi equation for geodesics can be guaranteed
if the metric admits a high enough number of commuting Killing vectors \cite{Benenti:1979}.

For the metric (\ref{met_all}), we introduce an orthonormal frame as
\beqa \label{frame}
&& {\bm e}^\mu = \frac{f_\mu}{\sqrt{P_\mu}}dx^{\mu} \,, \quad
   {\bm e}^{\hat{\mu}} = \sqrt{P_\mu}\sum_{k=0}^{n-1}A^{(k)}_{\mu} d\psi_k \,, \nonumber\\
&& {\bm e}^0 = \sqrt{S}\sum_{k=0}^n A^{(k)} d\psi_k \,. \label{tetrad_all}
\eeqa
Then, the rank-2 GCCKY tensor is given by
\beq
 {\bm h} = \sum_{\mu=1}^n x_\mu \,{\bm e}^\mu\wedge {\bm e}^{\hat{\mu}}
\eeq
with the torsion
\beqa
 {\bm T} &=& \sum_{\mu\neq\nu}\frac{2x_\mu\sqrt{P_\nu}(f_\mu-f_\nu)}{f_\mu f_\nu (x_\mu^2-x_\nu^2)}
           \,{\bm e}^\mu\wedge {\bm e}^{\hat{\mu}}\wedge {\bm e}^{\hat{\nu}} \nonumber\\
         & & + \varepsilon \sum_{\mu=1}^n \frac{2\sqrt{S}}{x_\mu}\left(\lambda-\frac{1}{f_\mu}\right)
           \,{\bm e}^\mu\wedge{\bm e}^{\hat{\mu}}\wedge{\bm e}^0 \,, \label{torsion_all}
\eeqa
where $\lambda$ is an arbitrary nonzero function which appears only in odd dimensions.
The ambiguity of $\lambda$ cannot be excluded by the GCCKY equation (\ref{GCCKYeq}),
e.g., see \cite{HKWY}.
It also has nothing to do with Einstein equations.
Even if we impose an Einstein equation, it determines the functions $\Xi_\mu$ and $f_\mu$,
but $\lambda$ is still arbitrary.

From the property that the wedge product of GCCKY tensors is a GCCKY tensor,
${\bm h}^{(j)}={\bm h}\wedge {\bm h}\cdots\wedge {\bm h}$ is a rank-$2j$ GCCKY tensor.
The Hodge dual ${\bm f}^{(j)}=* {\bm h}^{(j)}$ is a rank-$(D-2j)$ generalized KY tensor,
and its square, $K_{ab}^{(j)}=c_j f^{(j)}{}_{ac_1\cdots c_{D-2j-1}}f^{(j)}{}_b{}^{c_1\cdots c_{D-2j-1}}$
becomes a rank-2 KS tensor satisfying
$\nabla_{(a}K_{bc)} = 0$,
where $c_j$ is constant.
For an appropriate choice for $c_j$,
the KS tensors are written in the form
\beqa
 {\bm K}^{(j)} &=& \sum_{\mu=1}^n A^{(j)}_\mu ({\bm e}^\mu\otimes{\bm e}^\mu 
 + {\bm e}^{\hat{\mu}}\otimes{\bm e}^{\hat{\mu}}) \nonumber\\
  &+& \varepsilon A^{(j)}{\bm e}^0\otimes {\bm e}^0 \,, \label{KT_all}
\eeqa
where $j=0,1,\dots, n-1$. In particular, ${\bm K}^{(0)}={\bm g}^{(D)}$.
Thus, contracting with the tangent ${\bm p}=\dot{\gamma}$ to geodesics $\gamma$,
we obtain $n=[D/2]$ conserved quantities $\kappa^{(j)} = K^{(j)}{}^{ab}p_ap_b$ for $j=0,\dots,n-1$,
including the Hamiltonian $\kappa^{(0)}=H$.
In addition, since ${\bm \eta}^{(k)}={\bm \partial}_{\psi_k}$ for $k=0,\dots,n-1+\varepsilon$ are Killing vector fields,
we have $n+\varepsilon$ conserved quantities $\tilde{\kappa}^{(k)}=\eta^{(k)}{}^ap_a$.
The all conserved quantities $\{\kappa^{(j)},\tilde{\kappa}^{(k)}\}$ are in involution.

\subsection{Even dimensions}
In this section, we determine the unknown functions $\Xi_\mu(x_\mu)$
using the Einstein equations for perfect fluids in even dimensions.
The metric {\it ansatz} in $2n$ dimensions is given by
\begin{equation}
 {\bm g}^{(2n)} = \sum_{\mu=1}^n \frac{f_\mu^2}{P_\mu}dx_{\mu}^2
             + \sum_{\mu=1}^n  P_\mu \left( \sum_{k=0}^{n-1}A^{(k)}_{\mu} d\psi_k \right)^2 \,. \label{met_even}
\end{equation}
For the metric, we calculate the Ricci curvature (see Appendix B for details).
The off-diagonal components of the Ricci curvature are
\beqa
&& R_{\mu \hat{\mu}}=R_{\mu \nu}=R_{\mu \hat{\nu}}=0 \,, \nonumber\\
&& R_{\hat{\mu}\hat{\nu}}=\beta^2(D-2) \sqrt{P_\mu}\sqrt{P_\nu} \,. \label{ST0}
\eeqa
The diagonal components are
\beqa
 R_{\mu \mu} &=& I_\mu (P_T) +\beta^2\left[ I_\mu(P_T^{(2)}) 
                +\frac{3}{2}x_\mu \partial_\mu P_T + P_T\right] \,, \nonumber\\
 R_{\hat{\mu}\hat{\mu}} &=& R_{\mu \mu} + \beta^2(D-2) P_\mu \,, \label{ST0-2}
\eeqa
where
\beq
 P_T = \sum_{\mu=1}^n P_\mu \,, \quad
 P_T^{(2)} = \sum_{\mu=1}^n x_\mu^2 P_\mu
\eeq
and $I_\mu$ are differential operators given by
\beq
 I_\mu = -\frac{1}{2}\frac{\partial^2}{\partial x_\mu^2}
         +\frac{1}{x_\mu^2-x_\nu^2}\left(x_\mu\frac{\partial}{\partial x_\mu}
          -x_\nu\frac{\partial}{\partial x_\nu}\right) \,. \label{diffop}
\eeq

It should be emphasized that our metric ansatz is now expressed with a Euclidean signature,
so that we have to consider the Euclideanized Einstein equation for perfect fluids,
\begin{equation}
 R_{ab}-\frac{1}{2}R g_{ab}=-(\rho+p)u_a u_b+p g_{ab} \,, \label{EEE}
\end{equation}
where $u^au_a=1$. Eliminating the scalar curvature,
we obtain the Einstein equation in a convenient form
\beq
 R_{ab} = -(\rho+p)u_a u_b +\frac{\rho-p}{D-2} g_{ab} \,.
\eeq
Moreover, to solve the equation, we assume that perfect fluids are rigidly rotating; that is,
the velocity ${\bm u}$ is written as ${\bm u}=N{\bm \partial}_{\psi_0}$ where $N$ is the normalization function.
Since we have $N=1/\sqrt{P_T}$ from $u^au_a=1$,
the velocity is given in the canonical frame as
\begin{equation}
 {\bm u} = \frac{1}{\sqrt{P_T}} \sum_{\mu=1}^n \sqrt{P_\mu} {\bm e}_{\hat{\mu}} \,. \label{4vel}
\end{equation}
Under the assumption, together with (\ref{ST0}) and (\ref{ST0-2}),
the Einstein equation to solve reduces to
\beqa
 \frac{\rho-p}{D-2} &=& R_{11}=R_{22}=\cdots =R_{nn} \,, \label{EEE2} \\
 \frac{\rho+p}{D-2} &=& -\beta^2 P_T \,. \label{EEE3}
\eeqa

To solve Eq.~(\ref{EEE2}),
we notice that the $\mu\mu$ components of the Ricci curvature, $R_{\mu\mu}$,
can be written in a simple form.
Calculating $R_{\mu\mu}$ in terms of the functions $\Xi_\mu$
and their derivatives $\Xi_\mu^\prime$ and $\Xi_\mu^{\prime\prime}$, we obtain
\beqa
R_{\mu\mu} &=& -\frac{1}{2x_\mu}\left\{\frac{G_\mu}{U_\mu}
 -\sum_{\nu\neq\mu}\frac{2x_\mu}{x_\mu^2-x_\nu^2}\left[\frac{F_\mu}{U_\mu}+\frac{F_\nu}{U_\nu}\right]\right\} \nonumber\\
 & &-2\beta^2P_T \,,
\eeqa
where
\beqa
 G_\mu &=& x_\mu(1+\beta^2x_\mu^2)\Xi_\mu^{\prime\prime}+\beta^2x_\mu^2\Xi_\mu^\prime -4\beta^2x_\mu\Xi_\mu \,, \\
 F_\mu &=& x_\mu(1+\beta^2x_\mu^2)\Xi_\mu^\prime-(1+2\beta^2x_\mu^2)\Xi_\mu \,. \label{eq40}
\eeqa
Noticing that $G_\mu = F_\mu^\prime$ and that
\beq
\frac{\partial F_T}{\partial x_\mu} = \frac{F_\mu^\prime}{U_\mu}
 -\sum_{\nu\neq\mu}\frac{2x_\mu}{x_\mu^2-x_\nu^2}\left[\frac{F_\mu}{U_\mu}+\frac{F_\nu}{U_\nu}\right] \,,
\eeq
where
\beq
 F_T = \sum_{\rho=1}^n\frac{F_\rho}{U_\rho} \,, \label{FT}
\eeq
we obtain the following expressions for $R_{\mu\mu}$:
\beq
 R_{\mu\mu} = -\frac{1}{2x_\mu}\frac{\partial F_T}{\partial x_\mu}-2\beta^2 P_T \,. \label{rmm}
\eeq

Using the expression, $R_{\mu\mu}-R_{\nu\nu}=0$ implies that 
\beq
 \left[\frac{1}{x_\mu}\frac{\partial}{\partial x_\mu}
 -\frac{1}{x_\nu}\frac{\partial}{\partial x_\nu}\right] F_T = 0 \,.
\eeq
This can be solved by $F_T=F_T(\xi)$
where $F_T(\xi)$ is an arbitrary function of $\xi= \sum_{\mu=1}^nx_\mu^2$.
Substituting it into (\ref{FT})
and differentiating by $\partial_{x_1}\partial_{x_2}\cdots\partial_{x_n}$
both sides of the equation multiplied by the factor $\prod_{\mu\neq\nu}(x_\mu^2-x_\nu^2)$,
we arrive at the condition $F_T^{(n)}(\xi) = 0$,
which implies that $F_T(\xi)$ is an $(n-1)$th order polynomial in $\xi$.
Furthermore, going back to (\ref{FT}) again and comparing the coefficients of the equation,
we find that $F_T$ must be a linear function.
Namely, to be consistent with (\ref{FT}),
the function must be chosen as $F_T(\xi)=C_1 \xi + C_2$ where $C_1$ and $C_2$ are constants.
Then, using the identities
\beqa
&& \sum_{\mu=1}^n \frac{x_\mu^{2j}}{U_\mu} = 0 \,, \quad (j=0,\dots, n-2) \\
&& \sum_{\mu=1}^n \frac{x_\mu^{2(n-1)}}{U_\mu} = 1 \,, \quad
   \sum_{\mu=1}^n \frac{x_\mu^{2n}}{U_\mu} = \sum_{\mu=1}^n x_\mu^2 \,,
\eeqa
we obtain
\beq
 F_\mu = \sum_{k=0}^n c_{2k} x_\mu^{2k} \,, \label{EEE4}
\eeq
where $c_{2k}$ ($k=0,1,\dots, n$) are constants with $C_1=c_{2n}$ and $C_2=c_{2(n-1)}$.
In the end, using (\ref{eq40}) and (\ref{EEE4}), the problem of solving the Einstein equation (\ref{EEE})
has been reduced to that of solving first-order ordinary differential equations for $\Xi_\mu$,
\beq
 \Xi_\mu^\prime-\frac{1+2\beta^2x_\mu^2}{x_\mu(1+\beta^2x_\mu^2)}\Xi_\mu 
  - \frac{\sum_{k=0}^n c_{2k} x_\mu^{2k}}{x_\mu(1+\beta^2x_\mu^2)} = 0 \,.
\eeq
The general solution is
\beq
 \Xi_\mu = \sum_{k=0}^n c_{2k} \phi_{2k}(x_\mu)+a_\mu x_\mu \sqrt{1+\beta^2 x^2_\mu} \,, \label{EQ}
\eeq
where $a_\mu$ are integral constants, $\phi_0(x) \equiv -1$
and $\phi_{2k}(x)$ $(k=1,2,\dots)$ are given by
\beq
 \phi_{2k}(x) = x\sqrt{1+\beta^2x^2}\int_0^x \frac{t^{2(k-1)}dt}{(1+\beta^2t^2)^{3/2}} \,.
\eeq
Note that, for instance, we have
\beqa
 \phi_2(x)
&=& x^2 \,, \nonumber\\
 \phi_4(x)
&=& -\frac{x}{\beta^2}\left(x-\frac{\text{Arcsinh}(\beta x)\sqrt{1+\beta^2 x^2}}{\beta}\right) \,, \nonumber\\
 \phi_6(x)
&=& \frac{3x}{2\beta^4}\left(x+\frac{\beta^2}{3}x^3-\frac{\text{Arcsinh}(\beta x)\sqrt{1+\beta^2 x^2}}{\beta}\right) \,, \nonumber\\
  \phi_8(x)
&=& -\frac{15x}{8\beta^6}\left(x+\frac{\beta^2}{3}x^3-\frac{2\beta^4}{15}x^5\right. \\
& & \left.-\frac{\text{Arcsinh}(\beta x)\sqrt{1+\beta^2 x^2}}{\beta}\right) \,, \quad \dots \,. \nonumber
\eeqa
The solution contains parameters $c_{2k}$ ($k=0,\dots,n$), $a_\mu$ ($\mu=1,\dots,n$) and $\beta$.

In four dimensions, for $\mu=1,2$, we obtain
\begin{eqnarray}
 \Xi_\mu
&=& -c_0+c_2 x_\mu^2 +a_\mu x_\mu \sqrt{1+\beta^2 x^2_\mu} \\
& & -\frac{c_4x_\mu}{\beta^2}\left(x_\mu-\frac{\text{Arcsinh}(\beta x_\mu)\sqrt{1+\beta^2 x_\mu^2}}{\beta}\right) \,. \nonumber
\end{eqnarray}
The form coincides with the Wahlquist solution.

In the limit $\beta\to 0$, we have $\phi_{2k}\to x^{2k}/(2k-1)$.
The functions $\Xi_\mu$ take the form
\beq
 \Xi_\mu = \sum_{k=0}^n \tilde{c}_{2k} x_\mu^{2k} + a_\mu x_\mu \,,
\eeq
where $\tilde{c}_{2k}=c_{2k}/(2k-1)$.
This is the same form as Kerr-NUT-(A)dS metrics in $2n$ dimensions
found by Chen-L\"u-Pope \cite{Chen:2006}.

Finally, let us comment about the equation of state.
From (\ref{EEE2}), (\ref{EEE3}), (\ref{rmm}) and (\ref{EEE4}),
we have
\beq
 \frac{2\rho}{D-2} = -c_{2n}-3\beta^2 P_T \,, \quad
 \frac{2 p}{D-2} = c_{2n}+\beta^2 P_T \,. 
\eeq
Hence, the equation of state is $\rho+3p=(D-2)c_{2n}$.

\subsection{Odd dimensions}
Let us consider odd dimensions $D=2n+1$.
The metric {\it ansatz} in $2n+1$ dimensions is given by
\beqa
{\bm g}^{(2n+1)}
  &=& \sum_{\mu=1}^n \frac{f_\mu^2}{P_\mu}dx_{\mu}^2+\sum_{\mu=1}^n  P_\mu \left( \sum_{k=0}^{n-1}A^{(k)}_{\mu} d\psi_k \right)^2 \nonumber\\
  & & +S \left(\sum_{k=0}^n A^{(k)} d\psi_k \right)^2 \label{met_odd}
\eeqa
with unknown functions $\Xi_\mu$.
The off-diagonal components of the Ricci curvature are given by
\beqa
&& R_{\mu \nu}=R_{\mu \hat{\nu}}=R_{\mu \hat{\mu}}=R_{\mu 0}=0 \,, \nonumber\\
&& R_{\hat{\mu}\hat{\nu}}=\beta^2(D-2) \sqrt{P_\mu}\sqrt{P_\nu} \,, \label{ST1} \\
&& R_{\hat{\mu}0}=\beta^2(D-2) \sqrt{P_\mu}\sqrt{S} \,. \nonumber
\eeqa
The diagonal components are
\beqa
 R_{\mu \mu} &=& I_\mu (\tilde{P}_T) +\beta^2\left[ I_\mu(\tilde{P}_T^{(2)}) 
                +\frac{3}{2}x_\mu \partial_\mu \tilde{P}_T + \tilde{P}_T\right] \nonumber\\
             & & -\frac{1}{2x_\mu}\frac{\partial}{\partial x_\mu}
                 \left(\tilde{P}_T+\beta^2\tilde{P}^{(2)}_T\right) \,, \nonumber\\
 R_{\hat{\mu}\hat{\mu}} &=& R_{\mu \mu} + \beta^2(D-2) P_\mu \,, \label{ST0-2odd} \\
 R_{00} &=& -\sum_{\mu=1}^n \frac{1}{x_\mu}\frac{\partial}{\partial x_\mu}
                 \left(\tilde{P}_T+\beta^2\tilde{P}^{(2)}_T\right)+\beta^2\tilde{P}_T \nonumber\\
        & & +\beta^2(D-2)S \,, \nonumber
\eeqa
where
\begin{equation}
\tilde{P}_T=\sum_{\mu=1}^n \tilde{P}_\mu=P_T+S,~~~~\tilde{P}_T^{(2)}=\sum_{\mu=1}^n x_\mu^2 \tilde{P}_\mu=P_T^{(2)}
\end{equation}
and
\begin{equation}
\tilde{P_\mu}=\frac{\tilde{\Xi}_\mu}{U_\mu},~~~\tilde{\Xi}_\mu=\Xi_\mu-(-1)^n \frac{s_0^2}{x_\mu^2}.
\end{equation}

We assume that the velocity ${\bm u}$ lies in the plane of the Killing vectors
\begin{equation}
 {\bm u}=\frac{1}{\sqrt{\tilde{P}_T}} \left(\sum_{\mu=1}^n \sqrt{P_\mu} {\bm e}_{\hat{\mu}}+\sqrt{S} {\bm e}_0 \right).
\end{equation}
The equation reduces to
\beqa
 \frac{\rho-p}{D-2} &=& R_{11} = R_{22} = \cdots = R_{nn} \,, \label{EEE1_odd} \\
 \frac{\rho+p}{D-2} &=& -\beta^2\tilde{P}_T \,, \label{EEE2_odd}
\eeqa
and for all $\mu$,
\beq
 R_{00} = R_{\mu\mu} + \beta^2(D-2)S \,. \label{EEE3_odd}
\eeq

Similar to even dimensions, we find from the direct calculation that
the $\mu\mu$ and $00$ components of the Ricci curvature can be written in the simple form
\beqa
 R_{\mu\mu} &=& -\frac{1}{2x_\mu}\frac{\partial \tilde{F}_T}{\partial x_\mu}-2\beta^2 \tilde{P}_T \,, \label{rmm_odd} \\
 R_{00} &=& -\sum_{\mu=1}^n\frac{\tilde{F}_\mu}{x_\mu^2U_\mu}-2\beta^2\tilde{P}_T+\beta^2(D-2)S \,, \label{r00}
\eeqa
where
\beq
 \tilde{F}_T = \sum_{\mu=1}^n \frac{\tilde{F}_\mu}{U_\mu}
\eeq
and
\beq
 \tilde{F}_\mu = x_\mu(1+\beta^2x_\mu^2)\tilde{\Xi}_\mu^\prime-\beta^2x_\mu^2\tilde{\Xi}_\mu \,. \label{gmm1}
\eeq
As was discussed in even dimensions [cf.\ (\ref{EEE4})],
Eq.\ (\ref{EEE1_odd}) requires that $\tilde{F}_\mu$ take the form
\beq
 \tilde{F}_\mu = \sum_{k=0}^{n}c_{2k}x_\mu^{2k} \,. \label{gmm2}
\eeq
Indeed, by virtue of (\ref{rmm_odd}) and (\ref{r00}),
we easily see that (\ref{gmm2}) together with $c_0=0$ solves (\ref{EEE1_odd}) and (\ref{EEE3_odd}).
From the equality of (\ref{gmm1}) and (\ref{gmm2}), we obtain the first-order ordinary differential equations
\beq
 \tilde{\Xi}_\mu^\prime-\frac{\beta^2x_\mu}{1+\beta^2x_\mu^2}\tilde{\Xi}_\mu
 -\frac{\sum_{k=1}^{n}c_{2k}x_\mu^{2k-1}}{1+\beta^2x_\mu^2} = 0 \,.
\eeq
The general solution is
\begin{equation}
\tilde{\Xi}_\mu = \sum_{k=1}^{n} c_{2k}\tilde{\phi}_{2k}(x_\mu)+a_\mu \sqrt{1+\beta^2 x_\mu^2} \,,
\end{equation}
where $a_\mu$ are integral constants and $\tilde{\phi}_{2k}$ ($k=1,2,\dots$) are given by
\beq
 \tilde{\phi}_{2k}(x) = \sqrt{1+\beta^2x^2}\int_0^x \frac{t^{2k-1}dt}{(1+\beta^2t^2)^{3/2}} \,.
\eeq
Note that, for instance, we have
\begin{eqnarray}
 \tilde{\phi}_2(x)
&=& -\frac{1}{\beta^2}\left(1-\sqrt{1+\beta^2x^2}\right) \,, \nonumber\\
 \tilde{\phi}_4(x)
&=& \frac{2}{\beta^4}\left(1+\frac{\beta^2}{2}x^2-\sqrt{1+\beta^2x^2}\right) \,, \nonumber\\
 \tilde{\phi}_6(x)
&=& -\frac{8}{3\beta^6}\left(1+\frac{\beta^2}{2}x^2-\frac{\beta^4}{8}x^4-\sqrt{1+\beta^2x^2}\right) \,, \nonumber\\
 \tilde{\phi}_8(x)
&=& \frac{16}{5\beta^8}\left(1+\frac{\beta^2}{2}x^2-\frac{\beta^4}{8}x^4+\frac{\beta^6}{16}x^6\right. \\
& & \left.-\sqrt{1+\beta^2x^2}\right) \,, \quad \dots \,. \nonumber
\end{eqnarray}
Thus, we obtain
\beq
 \Xi_\mu = \sum_{k=1}^{n} c_{2k}\tilde{\phi}_{2k}(x_\mu)+a_\mu \sqrt{1+\beta^2 x_\mu^2}
 + \frac{(-1)^ns_0^2}{x_\mu^2} \,. \label{sol_odd}
\eeq
The solution contains parameters
$c_{2k}$ ($k=1,\dots,n$), $a_\mu$ ($\mu=1,\dots,n$), $k$ and $\beta$.

In five dimensions, for $\mu=1,2$, we have
\begin{eqnarray}
 \Xi_\mu
&=& -\frac{c_2}{\beta^2}\left(1-\sqrt{1+\beta^2x^2}\right)
    +a_\mu\sqrt{1+\beta^2x_\mu^2} + \frac{s_0^2}{x_\mu^2} \nonumber\\
& & +\frac{2c_4}{\beta^4}\left(1+\frac{\beta^2}{2}x^2-\sqrt{1+\beta^2x^2}\right) \,. \label{func_5d}
\end{eqnarray}

In the limit $\beta\to 0$, we have $\phi_{2k}\to x^{2k}/2k$.
The functions $\Xi_\mu$ take the form
\beq
 \Xi_\mu = \sum_{k=1}^{n} \tilde{c}_{2k}x_\mu^{2k}+a_\mu + \frac{(-1)^ns_0^2}{x_\mu^2} \,,
\eeq
where $\tilde{c}_{2k}=c_{2k}/2k$.
The form reproduces Kerr-NUT-(A)dS metrics in $2n+1$ dimensions \cite{Chen:2006}.

Since we have
\beq
 \frac{2\rho}{D-2} = -c_{2n}-3\beta^2 \tilde{P}_T \,, \quad
 \frac{2 p}{D-2} = c_{2n}+\beta^2 \tilde{P}_T \,,
\eeq
the equation of state is $\rho+3p=(D-2)c_{2n}$ like the even dimensional case.

\section{Separability}
We investigate the separability of the Hamilton-Jacobi for geodesics, Klein-Gordon and Dirac equations
for the higher-dimensional Wahlquist metrics (\ref{met_all}),
where we do not specify the functions $\Xi_\mu$ to deal with more general cases.
If we choose a particular form of the functions as (\ref{EQ}) in even dimensions
and (\ref{sol_odd}) in odd dimensions,
the results can be applied to those of the Wahlquist metrics.
In this section, we will see that the Hamilton-Jacobi for geodesics and Klein-Gordon equations
can be solved by separation of variables,
but we will not see the separation of variables in the Dirac equation in any dimension.
It is also shown that in even dimensions,
a modified Dirac equation with $1/3$ torsion \cite{HKWY2}
can be solved by separation of variables,
while it can not in odd dimensions.

\subsection{Separation of variables in the Hamilton-Jacobi equation for geodesics}
The separation of variables in Hamilton-Jacobi equations for geodesics occurs
if and only if the metric admits the separability structure established in \cite{Benenti:1979},
in which the corresponding Killing tensors can be written in the St\"ackel form
\beq
 K^{\mu \mu}_{(j)} = \bar{\phi}^{\mu}_{(j)} \,, \quad
 K^{k\ell}_{(j)} = \sum_{\mu=1}^n \zeta^{k\ell}_{(\mu)}\bar{\phi}^{\mu}_{(j)} \,,
\eeq
where $\bar{\phi}^{\mu}_{(j)}$ is the inverse matrix of the St\"ackel matrix
and $\zeta^{k \ell}_{(\mu)}$ are functions depending only on one variable $x_\mu$.

The Killing tensors ${\bm K}^{(j)}$ $(j=0,1, \dots, n-1)$ obtained in (\ref{KT_all}) are written 
in the coordinate basis as
\begin{eqnarray}
 {\bm K}^{(j)}
&=& \sum_{\mu=1}^{n}K_{(j)}^{\mu \mu}
    \,{\bm \partial}_{x_\mu} \otimes {\bm \partial}_{x_\mu}
    + \sum_{k,\ell=0}^{n-1+\varepsilon}K_{(j)}^{k\ell}
    \,{\bm \partial}_{\psi_k} \otimes {\bm \partial}_{\psi_\ell} \nonumber\\ 
&=& \sum_{\mu=1}^n \frac{A_{\mu}^{(j)}}{U_\mu}\Bigg[ \frac{\Xi_\mu}{f_\mu^2}
    \,{\bm \partial}_{x_\mu} \otimes {\bm \partial}_{x_\mu}
    + \frac{\varepsilon (-1)^{n+1}}{s_0^2 x_\mu^2}
    \,{\bm \partial}_{\psi_n} \otimes {\bm \partial}_{\psi_n} \nonumber\\
& & +\sum_{k,\ell=0}^{n-1+\varepsilon}\frac{(-1)^{k+\ell}x_\mu^{2(2n-2-k-\ell)}}{\Xi_\mu}
    \,{\bm \partial}_{\psi_k} \otimes {\bm \partial}_{\psi_\ell} \Bigg] \,. \nonumber
\end{eqnarray}
The St\"ackel matrix and the functions $\zeta^{k \ell}_{(\mu)}$ are given by
\beqa
 \phi^{(j)}_{\mu}
&=& \frac{(-1)^j x_\mu^{2(n-j-1)} f_\mu^2}{\Xi_\mu} \,, \label{ST} \\
 \zeta^{k \ell}_{(\mu)}
&=& \frac{(-1)^{k+\ell}f_{\mu}^2 x_\mu^{2(2n-2-k-\ell)}}{\Xi_\mu^2}
    + \frac{\varepsilon(-1)^{n+1}f_\mu^2}{s_0^2 x_\mu^2 \Xi_\mu}\delta_{k \ell}\delta_{k n} \,. \nonumber
\eeqa
In practice, the Hamiltonian-Jacobi equation for geodesics,
\begin{equation}
\frac{\partial S}{\partial \lambda}+g^{ab} \frac{\partial S}{\partial x^a}\frac{\partial S}{\partial x^b}=0 \,,
\end{equation}
allows an additive separation of variables
\begin{equation}
S=-\kappa_0 \lambda+\sum_{\mu=1}^n S_\mu(x_\mu)+\sum_{k=0}^{n-1+\varepsilon}n_k \psi_k,
\end{equation}
where $\kappa_0$ and $n_k$ are constants.
The functions $S_\mu$~($\mu=1,\dots,n)$ are given by
\begin{equation}
S_\mu(x_\mu)=\int \left( \sum_{j=0}^{n-1} \phi^{(j)}_\mu \kappa_i-
\sum_{k,\ell=0}^{n-1+\varepsilon} \zeta^{k\ell}_{(\mu)} n_k n_\ell \right)^{1/2} dx_\mu .
\end{equation}
In the limit $\beta\to 0$, this recovers the result for the Kerr-NUT-(A)dS metrics \cite{Frolov:2008,Yasui:2011}.

\subsection{Separation of variables in the Klein-Gordon equation}
The massive scalar field $\Phi$ is described by the Klein-Gordon equation
\begin{equation}
\frac{1}{\sqrt{g}}\frac{\partial}{\partial x^a}\sqrt{g}g^{ab} \frac{\partial \Phi}{\partial x^b}=m^2 \Phi
\end{equation}
This equation allows a multiplicative separation of variables
\begin{equation}
\Phi=\prod_{\mu=1}^n R_\mu(x_\mu) \prod_{k=0}^{n-1+\varepsilon}e^{in_k \psi_k},
\end{equation}
where the functions $R_{\mu} (x_\mu)$ $(\mu=1,\dots,n)$ satisfy the ordinary second order differential equations
\beqa
&&R_\mu^{\prime\prime}+
\left(\frac{\Xi_\mu^{'}}{\Xi_\mu}-\frac{f_\mu^{'}}{f_\mu}+\frac{\varepsilon}{x_\mu} \right)
 R_\mu^\prime \nonumber\\
&&+\left(\sum_{j=0}^{n-1} \phi^{(j)}_\mu \kappa_j-\sum_{k,\ell=0}^{n-1+\varepsilon} \zeta^{k\ell}_{(\mu)} n_k n_\ell \right)R_\mu=0
\eeqa
with $\kappa_0=-m^2$.
In the limit $\beta\to 0$, this recovers the result for the Kerr-NUT-(A)dS metrics \cite{Frolov:2008,Yasui:2011}.

\subsection{Separation of variables in the Dirac equation}
The existence of a GCCKY 2-form does not imply the separability of the Dirac equation.
However, we may expect the separability of a modified Dirac equation
which appears in the spacetimes admitting the GCCKY,
\begin{equation}\label{Dirac}
(\gamma^a D^{T/3}_a+m) \Psi=0,
\end{equation} 
where the Dirac operator has a 1/3 torsion, $T/3$, of the GCCKY 2-form, 
\begin{equation}
 D^{T/3}_a
= {\bm e}_a+\frac{1}{4} \gamma^b \gamma^c {\bm \omega}_{bc}({\bm e}_a)-\frac{1}{24}\gamma^b \gamma^c T_{abc} \,.
\end{equation}
The frame vector fields dual to (\ref{tetrad_all}) are given by
\beqa
&& {\bm e}_\mu=\frac{\sqrt{P_\mu}}{f_\mu}\frac{\partial}{\partial x_\mu} \,, \quad
   {\bm e}_0=\frac{\sqrt{S}}{s_0^2}\frac{\partial}{\partial \psi_n} \,, \nonumber\\
&& {\bm e}_{\hat{\mu}}=\sum_{k=0}^{n-1+\varepsilon}\frac{(-1)^k x_\mu^{2(n-1+k)}}{\sqrt{P_\mu} U_\mu}
   \frac{\partial}{\partial \psi_k} \,. \label{ortho_all_vec}
\eeqa
By using (\ref{connections}) and (\ref{torsion_all}),
the Dirac operator $\gamma^a D^{T/3}_a$ is explicitly written as
\beqa
&& \sum_{\mu=1}^n  \frac{\sqrt{P_\mu}}{f_\mu}\Bigg[
   \gamma^\mu \left(\frac{\partial}{\partial x_\mu}+\frac{\Xi^\prime_\mu}{4 \Xi_\mu}+\frac{\varepsilon}{2 x_\mu}
   +\frac{1}{2}\sum_{\nu \ne \mu}\frac{x_\mu}{x^2_\mu-x^2_\nu} \right) \nonumber\\
&& + \gamma^{\hat{\mu}} \left( \sum_{k=0}^{n-1+\varepsilon}
   \frac{(-1)^k x_\mu^{2(n-1+k)}f_\mu}{\Xi_\mu}\frac{\partial}{\partial \psi_k}
   +\sum_{\nu \ne \mu}\frac{x_\nu \gamma^\nu \gamma^{\hat{\nu}}}{2(x^2_\mu-x^2_\nu)} \right) \Bigg] \nonumber\\
&& + \varepsilon \gamma^0 \sqrt{S} \left[\frac{1}{s_0^2}\frac{\partial}{\partial \psi_n}-\frac{1}{4}\sum_{\mu=1}^n
   \left(\frac{\lambda}{x_\mu}+\frac{1}{f_\mu x_\mu}\right)\gamma^\mu \gamma^{\hat{\mu}} \right] \,.
\eeqa
The expression leads to the separation of variables in even dimensions.
The calculation is completely parallel to
that of \cite{Oota}.
We write the $2^n$ components of the spinor field as 
$\Psi_{\varepsilon_1\varepsilon_2\cdots\varepsilon_n} (\varepsilon_\mu=\pm 1)$.
Putting the separation solution
\begin{equation}
\Psi_{\varepsilon_1\varepsilon_2\cdots\varepsilon_n}
=\left(\prod_{1 \le \mu < \nu \le n}\frac{1}{\sqrt{x_\mu+\varepsilon_\mu \varepsilon_\nu x_\nu}} \right)
\hat{\Psi}_{\varepsilon_1\varepsilon_2\cdots\varepsilon_n} \,,
\end{equation}
where
\beq
 \hat{\Psi}_{\varepsilon_1\varepsilon_2\cdots\varepsilon_n}
= \left(\prod_{\mu=1}^n \chi^{(\mu)}_{\varepsilon_\mu}(x_\mu) \right) 
  \exp \left(i\sum_{k=0}^{n-1+\varepsilon}n_k \psi_k \right) \,
\eeq
the modified Dirac equation becomes
\beqa
&& \sum_{\mu=1}^n \frac{P^{(\mu)}_{\varepsilon_\mu}}{\prod_{\nu \ne \mu}
   (\varepsilon_\mu x_\mu-\varepsilon_\nu x_\nu)}+m \label{sep} \\
&& +\frac{i\varepsilon k}{\prod_{\rho=1}^n \varepsilon_\rho x_\rho}
   \left(\frac{n_n}{s_0^2}-\frac{1}{4}\sum_{\mu=1}^n\left( \frac{\lambda}{\varepsilon_\mu x_\mu}+
   \frac{1}{\varepsilon_\mu x_\mu f_\mu} \right) \right)=0 \,, \nonumber
\eeqa
where
\beqa \label{sep2}
 P^{(\mu)}_{\varepsilon_\mu}
&=& (-1)^{\mu-1}(\varepsilon_\mu)^{n-\mu}\frac{\sqrt{(-1)^{\mu-1}\Xi_\mu}}{f_\mu \chi^{(\mu)}_{\varepsilon_\mu}} \nonumber\\
& & \times \left(\frac{d}{dx_\mu}+\frac{\Xi^{'}_\mu}{4\Xi_\mu}+\frac{\varepsilon}{2 x_\mu}+\varepsilon_\mu Y_\mu \right)
    \chi^{(\mu)}_{-\varepsilon_\mu} \,, \\
 Y_\mu
&=& \sum_{k=0}^{n-1+\varepsilon}\frac{(-1)^k f_\mu x_\mu^{2(n-1+k)}}{\Xi_\mu}n_k \,.
\eeqa
Note that $P^{(\mu)}_{\varepsilon_\mu}$ depends only on the one variable $x_\mu$.

In even dimensions $(\varepsilon=0)$, Eq.\ (\ref{sep}) reduces to
\beq
 \sum_{\mu=1}^n \frac{P^{(\mu)}_{\varepsilon_\mu}}{\prod_{\nu \ne \mu}
 (\varepsilon_\mu x_\mu-\varepsilon_\nu x_\nu)}+m = 0 \,.
\eeq
The equation separates when
\begin{equation}\label{sep3}
 P^{(\mu)}_{\varepsilon_\mu}=\sum_{j=0}^{n-1}q_j (\varepsilon_\mu x_\mu)^j \,,
\end{equation}
where $q_{j}$ $(j=0, \dots, n-2)$ are arbitrary constants and $q_{n-1}=-m$.
Indeed, combining (\ref{sep2}) with (\ref{sep3}), we have the following coupled ordinary differential equations:
\beqa
&& \left(\frac{d}{dx_\mu}+\frac{\Xi^{'}_\mu}{4\Xi_\mu}
   +\frac{\varepsilon}{2 x_\mu}+\varepsilon_\mu Y_\mu \right)\chi^{(\mu)}_{-\varepsilon_\mu} \\
&& +\frac{(-1)^{\mu}(\varepsilon_\mu)^{n-\mu}f_\mu \sum_{j=0}^{n-1}q_j (\varepsilon_\mu x_\mu)^j}
   {\sqrt{(-1)^{\mu-1}\Xi_\mu}}\chi^{(\mu)}_{\varepsilon_\mu}=0 \,. \nonumber
\eeqa

In odd dimensions, we cannot separate (\ref{sep})
because of the last terms including the function $\lambda$.
For the Kerr-NUT-(A)dS metrics in odd dimensions, since we have $f_\mu=1$ for all $\mu$,
we are able to take $\lambda=1$ and then Eq.\ (\ref{sep}) can be solved by separation of variables \cite{Oota}.

\section{Summary and Discussion}
In Sec.~II, we have discovered a rank-2 GCCKY tensor with a skew-symmetric torsion
for the Wahlquist metric (\ref{Wahlquist})
which is a stationary, axially symmetric perfect fluid solution of the Einstein equation in four dimensions
with $\rho+3p=\text{const}$.
In Sec.~III, we have obtained stationary, axially symmetric perfect fluid solutions in higher dimensions,
where we have made use of canonical forms of metrics admitting a rank-2 GCCKY tensor and
have directly solved the higher-dimensional Einstein equations in higher dimensions.
The exact solutions obtained generalize the Wahlquist metric in four dimensions to all even dimensions (\ref{met_even})
and odd ones (\ref{met_odd}),
for which the equations of state are always given by $\rho+3p=\text{const}$.
As far as we know, they are the first examples of rotating perfect fluid solutions in higher dimensions.

We could solve the Einstein equations for perfect fluids due to the presence of a rank-2 GCCKY tensor.
In this sense, if we find another solution admitting a rank-2 GCCKY tensor in four dimensions,
it might be possible to generalize it to higher dimensions.
When we solved the Einstein equations, as an {\it ansatz},
we have focused on type A metrics admitting a rank-2 GCCKY tensor,
but it would be of great interest to investigate the other types of metrics (called type B and type C \cite{HKWY}).
Since we have assumed a particular case of type A metrics,
it might be possible to find other perfect fluid solutions of type A even in four dimensions.
Another thing we assumed is that perfect fluids are rigidly rotating,
so it would be worth asking whether the assumption can be relaxed.

In Sec.~IV, we have investigated the separability of the Hamilton-Jacobi for geodesics, Klein-Gordon
and (both standard and modified) Dirac equations 
for the obtained higher-dimensional perfect fluid solutions.
In four dimensions, the Wahlquist metric shares the similar separability to the Kerr metric.
The Hamilton-Jacobi for geodesics, Klein-Gordon and modified Dirac equation with a $1/3$ torsion
equations can be solved by separation of variables.
We have seen that
the Hamilton-Jacobi for geodesics and Klein-Gordon equations separate also in higher dimensions,
which is responsible for the rank-2 GCCKY tensor.
Although the Dirac equation does not separate in arbitrary dimensions,
the modified Dirac with $1/3$ torsion equation does only in even dimensions.
In odd dimensions, there is an obstruction.
Even for any choice of the function $\lambda$ in (\ref{torsion_all}),
the modified Dirac equation does not separate.

Since the equations of state are given by $\rho+3p=\text{const}$,
the present situation seems to be unrealistic for compressible fluids.
Even so, it would be interesting to consider whether the obtained solutions describe the interiors of rotating stars
and (not necessarily smoothly) connect to vacuum spacetimes.
For instance, when we take the static limit of the Wahlquist solution,
we obtain static, spherically symmetric perfect fluid solution (\ref{Whit}).
Then, it is possible to match the metric to the Schwarzschild metric,
as was discussed in 4 \cite{Wahlquist:1968,Whittaker:1968}.
In higher dimensions, for static, spherically symmetric perfect fluid solutions with $\rho=\text{const.}$ \cite{Krori:1988,Zarro:2009}
and $p=-\rho$ \cite{Shen:1989}, the similar matching conditions were discussed,
where the metrics are joined to the Schwarzschild-Tangherlini metrics in arbitrary dimensions.

\section*{Acknowledgments}
We would like to thank Hideki Maeda for introducing us to this project.
We would also like to thank David Kubiz\v{n}\'ak for reading the draft carefully and giving us useful comments.
We are grateful to Masashi Kimura, Shunichiro Kinoshita and Masato Nozawa for the helpful discussion,
and also grateful to Gary W.\ Gibbons for reading the draft.
T.H. was supported by Research Center for Measurement in Advanced Science (RCMAS), Rikkyo University.
Y.Y. was supported by the Grant-in-Aid for Scientific Research No. 23540317.

\appendix

\section{Particular limits of the Wahlquist spacetime in four dimensions}
It is known that as the particular limits,
many known solutions can be included in the Wahlquist metric in four dimensions
\cite{Wahlquist:1968,Kramer:1985,Senovilla:1987,Wahlquist:1992,Mars:2001}.
We review the relationship here, again.
A detailed explanation can be found also in \cite{Mars:2008}.
The metric (\ref{Wah}) is written in a local coordinate system $(r,p,\tau,\sigma)$ as
\beqa
ds^2 &=& -\frac{{\cal Q}}{r^2+p^2}(d\tau+p^2 d\sigma)^2+\frac{{\cal P}}{r^2+p^2}(d\tau-r^2 d\sigma)^2 \nonumber\\
     & & +\frac{r^2+p^2}{{\cal Q}(1-\beta^2 r^2)}dr^2 +\frac{r^2+p^2}{{\cal P}(1+\beta^2 p^2)}dp^2 \label{Wah_app1}
\eeqa
with the functions
\beqa
 {\cal Q} &=& Q_0 + a_2 r\sqrt{1-\beta^2r^2} + \nu_0 r^2 \nonumber\\
   &&    + \frac{\mu_0}{\beta^2}\left[r^2-\frac{r \text{Arcsin}(\beta r)\sqrt{1-\beta^2r^2}}{\beta}\right] \,, \\
 {\cal P} &=& Q_0 + a_1 p\sqrt{1+\beta^2p^2} - \nu_0 p^2 \nonumber\\
   &&    - \frac{\mu_0}{\beta^2}\left[p^2-\frac{p \text{Arcsinh}(\beta p) \sqrt{1+\beta^2p^2}}{\beta}\right] \,.
\eeqa
The parameters are $Q_0$, $\nu_0$, $a_1$, $a_2$, $\mu_0$ and $\beta$.
Of them, the only five parameters are independent,
which correspond to mass, rotation, NUT, cosmological constant, the perfect fuild's parameters.

\subsection{Kerr-NUT-(A)dS limit}
Let us consider the metric (\ref{Wah_app1}) with $\beta=0$.
In the limit $\beta\to 0$, we have
\beqa
 \text{Arcsin}(\beta r)\sqrt{1-\beta^2r^2} &=& \beta r - \frac{\beta^3r^3}{3} + \cdots \,, \nonumber\\
 \text{Arcsinh}(\beta p)\sqrt{1+\beta^2p^2} &=& \beta p  + \frac{\beta^3 p^3}{3} + \cdots \,. \nonumber
\eeqa
Hence, the metric becomes
\beqa
ds^2 &=& -\frac{{\cal Q}}{r^2+p^2}(d\tau+p^2 d\sigma)^2+\frac{{\cal P}}{r^2+p^2}(d\tau-r^2 d\sigma)^2 \nonumber\\
     & & +\frac{r^2+p^2}{{\cal Q}}dr^2 +\frac{r^2+p^2}{{\cal P}}dp^2 \,,
\eeqa
where
\beqa
 {\cal Q} &=& Q_0 + a_2 r + \nu_0 r^2 - \frac{\mu_0}{3}r^4 \,, \\
 {\cal P} &=& Q_0 + a_1 p - \nu_0 p^2 - \frac{\mu_0}{3}p^4 \,.
\eeqa
The form was investigated first by Carter \cite{Carter:1968}, which is called the Kerr-NUT-(A)dS metric.
This is also a particular case of Plebanski \cite{Plebanski:1975} and Plebanski-Demianski \cite{Plebanski:1976} metrics.
Recently, the geometric characterization was investigated in \cite{Mars:2013}.
For the physical meaning of the parameters, e.g., see \cite{GP:2007}.

\subsection{Static limit}
It is possible to take the static limit of the Wahlquist metric (\ref{Wah_app1}),
as was pointed out in \cite{Wahlquist:1968,Mars:2001}.
If we perform the coordinate transformation $p= a\cos\theta$, $\tau = t - a\phi$ and $\sigma = \phi /a$
and then send $a\to 0$ (with $Q_0=a^2\tilde{Q_0}$ and $a_1=a\tilde{a}_1$), the metric becomes
\beq
ds^2 = -f(r)dt^2 +\frac{dr^2}{(1-\beta^2r^2)f(r)}+r^2d\Omega^2 \,,
\eeq
where $d\Omega^2$ is a two-dimensional metric with the constant curvature $\nu_0$ and
\beqa
 f(r) &=& \nu_0+\frac{a_2\sqrt{1-\beta^2 r^2}}{r} \nonumber\\
      & & +\frac{\mu_0}{\beta^2}\left[1-\frac{\text{Arcsin}(\beta r)\sqrt{1-\beta^2 r}}{\beta r}\right] \,. \label{Whit}
\eeqa
If we take $\nu_0>0$, it is the Whittaker metric \cite{Whittaker:1968}
which is a static, spherically symmetric perfect fluid solution
describing an interior of Schwarzschild spacetime.

\section{Special cases of the Wahlquist metric in five dimensions}
In this appendix, we would like to discuss special cases of the higher-dimensional Wahlquist metrics
obtained in Sec.~III, especially in five dimensions.
Before doing so, for the metric (\ref{met_odd}) with the functions (\ref{func_5d}) in five dimensions,
we perform the Wick rotation $x=ir$. Then, the metric is written as
\beqa
 ds^2_5
 &=& \frac{r^2+y^2}{(1-\beta^2r^2)\Xi_1}dr^2+\frac{r^2+y^2}{(1+\beta^2y^2)\Xi_2}dy^2 \label{met_5D}\\
 & & - \frac{\Xi_1}{r^2+y^2}{\bm w}_1^2 + \frac{\Xi_2}{r^2+y^2}{\bm w}_2^2
     - \frac{s_0^2}{r^2y^2}{\bm w}_3^2 \,, \nonumber
\eeqa
where
\beqa
&& {\bm w}_1 = d\psi_0 + y^2 d\psi_1 \,, \quad
   {\bm w}_2 = d\psi_0 - r^2 d\psi_1 \,, \nonumber\\
&& {\bm w}_3 = d\psi_0 + (y^2-r^2)d\psi_1 - r^2y^2 d\psi_2 \,.
\eeqa
The functions $\Xi_\mu$ are given by
\beqa
&& \Xi_1 = \frac{2c_4}{\beta^4}\left(1-\frac{\beta^2}{2}r^2-\sqrt{1-\beta^2r^2}\right)
           -\frac{s_0^2}{r^2} \nonumber\\
&&  ~~~~~~~~~~ -\frac{c_2}{\beta^2}\left(1-\sqrt{1-\beta^2r^2}\right) + a_1\sqrt{1-\beta^2r^2} \,, \nonumber\\
&& \Xi_2 = \frac{2c_4}{\beta^4}\left(1+\frac{\beta^2}{2}y^2-\sqrt{1+\beta^2y^2}\right)
           +\frac{s_0^2}{y^2} \\
&&  ~~~~~~~~~~ -\frac{c_2}{\beta^2}\left(1-\sqrt{1+\beta^2y^2}\right) + a_2\sqrt{1+\beta^2y^2} \,. \nonumber
\eeqa

\subsection{Kerr-(A)dS metric}
Taking $\beta=0$ leads to the vacuum solution,
in which the energy density and pressure of the perfect fluid vanish
and the metric takes the form obtained by Chen-L\"u-Pope \cite{Chen:2006},
which was previously obtained in \cite{Hawking:1999,Gibbons:2004}.

\subsection{Rotating perfect fluids with equal angular momenta}
It is shown \cite{Gibbons:2004} that when all angular momenta are set to equal in odd dimensions $D=2n+1$,
Myers-Perry-(A)dS metrics can be recast in a simpler form,
in which the Hopf fibrations over $\mathbb{CP}^{n-1}$ appear in the metrics.
For instance, the Hopf fibration over $\mathbb{CP}^1 \cong S^2$ appears in 5 dimensions.
It is realized from the viewpoint of Killing-Yano symmetry that
the eigenvalues of Killing-Yano tensors change from functions to constants (from nondegenerate to degenerate).
Actually, if we start with a metric {\it ansatz} admitting a degenerate CCKY 2-form,
we obtain a vacuum solution including the Myers-Perry-(A)dS metrics with equal angular momenta \cite{Houri:2008b,Yasui:2011}.
In analogy with it, it is possible to consider a metric {\it ansatz} admitting a degenerate GCCKY 2-form
and then, as is expected, we obtain a stationary, axially symmetric perfect fluid solution.
The metric in five dimensions is written in a local coordinate system ($t,r,\theta,\phi,\psi$) as
\beqa
 ds^2_5
      &=& -f(r)\left(dt+ \frac{a}{2\lambda} {\bm \sigma}_3 \right)^2
          +\frac{1+\beta^2a^2}{(1-\beta^2r^2)f(r)}dr^2 \nonumber\\
      & & +\frac{k^2}{r^2}\Big(a dt +\frac{r^2+a^2}{2\lambda} {\bm \sigma}_3 \Big)^2
          +\frac{r^2+a^2}{4\lambda}d\Omega^2_{S^2} \,, \nonumber\\
      \label{met_5d_equal}
\eeqa
where $d\Omega^2_{S^2}$ is the standard metric on the two-dimensional unit sphere $S^2$,
and ${\bm \sigma}_3$ is the 1-form such that $d{\bm \sigma}_3$ is the K\"ahler form of $d\Omega^2_{S^2}$.
The function $f(r)$ is given by
\beqa
 f(r) &=& \frac{1}{r^2+a^2}\Bigg[k^2a^2\left(\frac{a^2}{r^2}+3\right)
          -2M\sqrt{1-\beta^2r^2} \nonumber\\
      & & +\frac{2k^2+c_2a^2}{\beta^2}\left(1-\sqrt{1-\beta^2r^2}\right) \nonumber\\
      & & +\frac{2c_2}{\beta^4}\left(1-\frac{\beta^2}{2}r^2-\sqrt{1-\beta^2r^2}\right)\Bigg] \,,
\eeqa
where
\beq
 \lambda = \frac{(4+3\beta^2a^2)k^2-c_2a^2}{4(1+\beta^2a^2)} \,.
\eeq
The metric contains five parameters $M$, $a$, $c_2$, $k$ and $\beta$.
The velocity of the perfect fluid is ${\bm u}=1/\sqrt{-g_{tt}}\,{\bm \partial}_t$.
The enegery density and pressure are
\beq
 \rho = -\frac{3(3\beta^2 g_{tt}+c_2)}{2(1+\beta^2a^2)} \,, \quad
 p = \frac{3(\beta^2g_{tt}+c_2)}{2(1+\beta^2a^2)} \,,
\eeq
where $g_{tt}=-f(r)+k^2a^2/r^2$.
The equation of state is given by
\beq
 \rho + 3p = \frac{3c_2}{1+\beta^2a^2} \,.
\eeq

\subsection{Static limit}
The static limit of the metric (\ref{met_5d_equal}) is given by $a=0$.
Then, since we have $k^2=\lambda$, the metric is written as
\beqa
 ds^2_5 = - f(r)dt^2 +\frac{dr^2}{(1-\beta^2r^2)f(r)}
           + \frac{r^2}{\lambda} d\Omega^2_3 \,,
\eeqa
where $d\Omega_3^2$ is the standard metric on the unit sphere $S^3$
and the function $f(r)$ is given by
\beqa
 f(r) &=& \frac{2\lambda}{\beta^2r^2}\left(1-\sqrt{1-\beta^2r^2}\right)
          -\frac{2M\sqrt{1-\beta^2r^2}}{r^2} \nonumber\\
      & & +\frac{2c_2}{\beta^4r^2}\left(1-\frac{\beta^2}{2}r^2-\sqrt{1-\beta^2r^2}\right) \,.
\eeqa
As a consequence, the static metric becomes spherically symmetric.

\section{Curvature quantities of the higher-dimensional Wahlquist metrics}
By using the tetrad method, this appendix calculates the curvature quantities
for the tetrad (\ref{frame}) of the higher-dimensional metrics (\ref{met_all}).
In what follows, the indices $\mu$ and $\nu$ are different and no sum.

From the first structure equation $d{\bm e}^a + {\bm \omega}^a{}_b \wedge {\bm e}^b = 0$
and ${\bm \omega}_{ba} = -{\bm \omega}_{ab}$,
the connection 1-forms are calculated as follows:
\begin{eqnarray}
 {\bm \omega}^\mu{}_\nu
&=& -\frac{x_\nu \sqrt{P_\nu}}{f_\nu(x_\mu^2-x_\nu^2)}{\bm e}^\mu
    -\frac{x_\mu \sqrt{P_\mu}}{f_\mu(x_\mu^2-x_\nu^2)}{\bm e}^\nu \,, \nonumber\\
 {\bm \omega}^\mu{}_{\hat{\mu}}
&=& -\frac{\partial_\mu \sqrt{P_\mu}}{f_\mu}{\bm e}^{\hat{\mu}}
    +\sum_{\nu \ne \mu}\frac{x_\mu \sqrt{P_\nu}}{f_\mu(x_\mu^2-x_\nu^2)}{\bm e}^{\hat{\nu}} \nonumber\\
& & +\frac{\varepsilon \sqrt{S}}{f_\mu x_\mu}{\bm e}^0 \,, \nonumber\\
 {\bm \omega}^\mu{}_{\hat{\nu}}
&=& \frac{x_\mu \sqrt{P_\nu}}{f_\mu(x_\mu^2-x_\nu^2)}{\bm e}^{\hat{\mu}}
    -\frac{x_\mu \sqrt{P_\mu}}{f_\mu(x_\mu^2-x_\nu^2)}{\bm e}^{\hat{\nu}} \,,\nonumber \\
 {\bm \omega}^{\hat{\mu}}{}_{\hat{\nu}}
&=& -\frac{x_\mu \sqrt{P_\nu}}{f_\mu(x_\mu^2-x_\nu^2)}{\bm e}^{\mu}
    -\frac{x_\nu \sqrt{P_\mu}}{f_\nu(x_\mu^2-x_\nu^2)}{\bm e}^{\nu} \,,\nonumber \\
 {\bm \omega}^\mu{}_0
&=& \frac{\sqrt{S}}{f_\mu x_\mu}{\bm e}^{\hat{\mu}}
    -\frac{\sqrt{P_\mu}}{f_\mu x_\mu}{\bm e}^0 \,,\nonumber\\
 {\bm \omega}^{\hat{\mu}}{}_0
&=& -\frac{\sqrt{S}}{f_\mu x_\mu}{\bm e}^{\mu} \,. \label{connections}
\end{eqnarray}

From the second structure equation ${\bm R}^a{}_b = d{\bm \omega}^a{}_b+{\bm \omega}^a{}_c\wedge{\bm \omega}^c{}_b$,
the curvature 2-forms are calculated as follows:
\begin{eqnarray}
 {\bm R}^\mu{}_\nu
&=& R^\mu{}_{\nu \mu \nu} \,{\bm e}^\mu\wedge{\bm e}^\nu
    +R^\mu{}_{\nu \hat{\mu} \hat{\nu}} \,{\bm e}^{\hat{\mu}} \wedge {\bm e}^{\hat{\nu}}, \nonumber\\
 {\bm R}^\mu{}_{\hat{\mu}}
&=& R^\mu{}_{\hat{\mu} \mu \hat{\mu}}\,{\bm e}^{\mu} \wedge {\bm e}^{\hat{\mu}}
    +\sum_{\nu \ne \mu}R^\mu{}_{\hat{\mu} \nu \hat{\nu}}\,{\bm e}^\nu \wedge {\bm e}^{\hat{\nu}} \nonumber\\
& & +\beta^2 \sum_{\nu \ne \mu}\sqrt{P_\mu}\sqrt{P_\nu}\,{\bm e}^\mu \wedge {\bm e}^{\hat{\nu}} \nonumber\\
& & +\varepsilon \beta^2 \sqrt{P_\mu} \sqrt{S} \,{\bm e}^\mu \wedge {\bm e}^0, \nonumber\\
 {\bm R}^\mu{}_{\hat{\nu}}
&=& R^\mu{}_{\hat{\nu} \mu \hat{\nu}}\,{\bm e}^\mu \wedge {\bm e}^{\hat{\nu}}
    +R^\mu{}_{\hat{\nu} \nu \hat{\mu}}\,{\bm e}^\nu \wedge {\bm e}^{\hat{\mu}} \nonumber\\
& & +\beta^2 \sqrt{P_\mu}\sqrt{P_\nu}\,{\bm e}^\mu \wedge {\bm e}^{\hat{\mu}}\nonumber\\
& & +\beta^2 \sum_{\rho \ne \mu, \nu}\sqrt{P_\nu}\sqrt{P_\rho}\,{\bm e}^\mu \wedge {\bm e}^{\hat{\rho}} \nonumber\\
& & +\varepsilon \beta^2 \sqrt{P_\nu} \sqrt{S}\, {\bm e}^\mu \wedge {\bm e}^0,\nonumber\\
 {\bm R}^{\hat{\mu}}{}_{\hat{\nu}}
&=& R^{\hat{\mu}}{}_{ \hat{\nu} \mu \nu}\,{\bm e}^\mu \wedge {\bm e}^{\nu}
    +R^{\hat{\mu}}{}_{\hat{\nu} \hat{\mu} \hat{\nu}}\,{\bm e}^{\hat{\mu}} \wedge {\bm e}^{\hat{\nu}} \nonumber\\
& & +\beta^2 \sum_{\rho \ne \mu, \nu} \sqrt{P_\nu}\sqrt{P_\rho}\,{\bm e}^{\hat{\mu}} \wedge {\bm e}^{\hat{\rho}}\nonumber\\
& & -\beta^2 \sum_{\rho \ne \mu, \nu}\sqrt{P_\mu}\sqrt{P_\rho}\,{\bm e}^{\hat{\nu}} \wedge {\bm e}^{\hat{\rho}} \nonumber\\
& & +\varepsilon \beta^2 \sqrt{P_\nu} \sqrt{S}\, {\bm e}^{\hat{\mu}} \wedge {\bm e}^0 \nonumber\\
& & -\varepsilon \beta^2 \sqrt{P_\mu} \sqrt{S} \,{\bm e}^{\hat{\nu}} \wedge {\bm e}^0,\nonumber\\
 {\bm R}^\mu{}_0
&=& R^\mu{}_{0 \mu 0}\,{\bm e}^\mu \wedge {\bm e}^0
    +\beta^2 \sqrt{P_\mu} \sqrt{S} \,{\bm e}^\mu \wedge {\bm e}^{\hat{\mu}} \nonumber\\
& & +\beta^2\sum_{\nu \ne \mu} \sqrt{P_\nu} \sqrt{S} \,{\bm e}^\mu \wedge {\bm e}^{\hat{\nu}}, \nonumber\\
 {\bm R}^{\hat{\mu}}{}_0
&=& R^{\hat{\mu}}{}_{0 \hat{\mu} 0}\,{\bm e}^{\hat{\mu}} \wedge {\bm e}^0
    +\beta^2 \sum_{\nu \ne \mu}\sqrt{P_\nu} \sqrt{S} \,{\bm e}^{\hat{\mu}} \wedge{\bm e}^{\hat{\nu}} \nonumber\\
& & +\beta^2\sum_{\nu \ne \mu} \sqrt{P_\mu} \sqrt{P_\nu}\,{\bm e}^{\hat{\nu}} \wedge {\bm e}^0,
\end{eqnarray}
The components of the curvature 2-forms are given by
\begin{eqnarray}
 R^\mu{}_{\nu \mu \nu}
&=& -\frac{1}{2(x_\mu^2-x_\nu^2)}\left( x_\mu \partial_\mu \tilde{P}_T-x_\nu \partial_\nu \tilde{P}_T \right), \nonumber\\
& & -\frac{\beta^2}{2(x_\mu^2-x_\nu^2)}\left( x_\mu \partial_\mu \tilde{P}^{(2)}_T-x_\nu \partial_\nu \tilde{P}^{(2)}_T \right), \nonumber\\
 R^\mu{}_{\nu \hat{\mu} \hat{\nu}}
&=& -\frac{1}{2f_\mu f_\nu(x_\mu^2-x_\nu^2)}\left( x_\nu \partial_\mu \tilde{P}_T-x_\mu \partial_\nu \tilde{P}_T \right), \nonumber\\
 R^\mu{}_{\hat{\mu} \mu \hat{\mu}}
&=& -\frac{1}{2}\partial^2_{\mu}\tilde{P}_T-\frac{\beta^2}{2}\partial^2_\mu\tilde{P}^{(2)}_T
    +\frac{3 \beta^2}{2}x_\mu \partial_\mu \tilde{P}_T+\beta^2 P_\mu, \nonumber\\
R^\mu{}_{0 \mu 0}&=&-\frac{1}{2 x_\mu}(\partial_\mu \tilde{P}_T+\beta^2 \partial_\mu \tilde{P}^{(2)}_T)+\beta^2 S
\end{eqnarray}
and
\begin{eqnarray}
&& R_{\hat{\mu} \nu \hat{\mu} \nu} = R_{\mu \nu \mu \nu} +\beta^2 P_\mu \,, \nonumber\\
&& R_{\hat{\mu} \hat{\nu} \hat{\mu} \hat{\nu}} = R_{\mu \nu \mu \nu} + \beta^2 (P_\mu+P_\nu) \,, \nonumber\\
&& R_{\mu \hat{\mu} \nu \hat{\nu}} = 2 R_{\mu \nu \hat{\mu} \hat{\nu}} \,, \quad
   R_{\mu \hat{\nu} \nu \hat{\mu}} = R_{\mu \nu \hat{\mu} \hat{\nu}} \,, \\
&& R_{\hat{\mu} 0 \hat{\mu} 0} = R_{\mu 0 \mu 0} + \beta^2 P_\mu \,. \nonumber
\end{eqnarray}
The functions $\tilde{P}_T$ and $\tilde{P}^{(2)}_T$ are defined by
\begin{eqnarray}
\tilde{P}_T &=&\sum_{\mu=1}^n P_\mu+\varepsilon S,~~\tilde{P}^{(2)}_T=\sum_{\mu=1}^n x_\mu^2 P_\mu.
\end{eqnarray}

In \cite{Houri:2007a}, it was shown that the higher-dimensional Kerr-NUT-(A)dS metrics are of type D in all dimensions.
This motivates us to ask if the higher-dimensional Wahlquist metrics obtained in Sec.~III are also of type D,
because the Kerr-NUT-(A)dS metrics are obtained as the limit of the Wahlquist metrics.
To see it, since we need to prepare a null orthonormal frame $\{ {\bm k}, {\bm l}, {\bm e}_\alpha \}$
such that ${\bm k}$ is a Weyl aligned null direction,
we define it for the higher-dimensional Wahlquist metrics 
in the way similar to higher-dimensional Kerr-NUT-(A)dS metrics.
Using the orthonormal frame (\ref{ortho_all_vec}), for a fixed number $\mu$, we define
\beqa
 {\bm k} &=& \frac{1}{\sqrt{2 P_\mu}}({\bm e}_\mu+\sqrt{-1}{\bm e}_{\hat{\mu}}) \,, \\
 {\bm l} &=& \frac{\sqrt{P_\mu}}{\sqrt{2}}({\bm e}_\mu-\sqrt{-1}{\bm e}_{\hat{\mu}}) \,.
\eeqa
By definition, this frame yields $(\alpha \ne \beta)$
\beqa
&& ({\bm k}, {\bm k})=({\bm l}, {\bm l})=0 \,, \quad ({\bm k}, {\bm l})=1 \,, \nonumber\\
&& ({\bm k}, {\bm e}_\alpha)=({\bm l}, {\bm e}_\alpha)=({\bm e}_\alpha, {\bm e}_\beta)=0 \,.
\eeqa
We also have $\nabla_{k} {\bm k}=0$,
which means that the integral curve of ${\bm k}$ is a geodesic.
It is easy to see that the Weyl curvature yields the type D condition \cite{Coley:2004}
\begin{eqnarray}
&& W({\bm k}, {\bm e}_\alpha, {\bm e}_\beta, {\bm e}_\gamma) = W({\bm l}, {\bm e}_\alpha, {\bm e}_\beta, {\bm e}_\gamma)=0 \,, \nonumber\\
&& W({\bm k}, {\bm e}_\alpha, {\bm k}, {\bm e}_\alpha) =  W({\bm l}, {\bm e}_\alpha, {\bm l}, {\bm e}_\alpha)=0 \,, \nonumber\\
&& W({\bm k}, {\bm l}, {\bm k}, {\bm e}_\alpha) = W({\bm k}, {\bm l}, {\bm l}, {\bm e}_\alpha)=0 \,.
\end{eqnarray}
We thus find that the higher-dimensional Wahlquist metrics
are of type D in all dimensions.


\end{document}